# Investigation of low-lying energy spectra for deformed prolate nuclei via partial dynamical $SU(3)$ symmetry

N. Fouladi[a], J. Fouladi[a] , H. Sabri[a1]

[a] Faculty of Physics, University of Tabriz, Tabriz 51664, Iran.

---

[1] E-mail: h-sabri@tabrizu.ac.ir

**Abstract**

We consider the possibility of identifying nuclei exhibiting the partial dynamical $SU(3)$ symmetry ($SU(3)$-PDS) as those having excitation energy ratio $R_{4/2} \geq 3.00$. For this purpose, the level energy spectra of a set of 51 nuclei in the rare earth and actinide regions which presenting an axially deformed prolate rotational structure were analyzed via nuclear partial dynamical $SU(3)$ symmetry in the framework of interacting boson model, to see if the $SU(3)$-PDS is broadly applicable, and where, how, and in which nuclei it breaks down. Overall, the PDS works very well, the predictions of such intermediate symmetry structure for energy spectrum were compared with the most recent experimental data and an acceptable degree of agreement is achieved. We conclude that PDS predictions have a more regular behavior in description of axially deformed prolate rotational nuclei than DS, which may lead to accurate predictions of such nuclei, and hence play a significant role in understanding the regular behavior of complex nuclei.

## I. Introduction

Dynamical symmetries (DS) have attracted extensive interest in recent decades due to their unprecedented power to descriptions of the complex systems properties, which have led to many important discoveries in diverse areas of physics with notable examples in nuclear, molecular, hadronic, polymer and nanostructure physics [1-5].Dynamical symmetries provide considerable insight into the nature of complex systems which can be viewed as a generalization and refinement of the exact symmetry concepts. In a dynamical symmetry, the Hamiltonian commutes with the Casimir operator of the symmetry group (G), i.e. $[\hat{H}, \hat{C}_G] = 0$. It means, the block structure of H is retained, the states preserve the good symmetry but, in general, are no longer degenerate. When the symmetry is completely broken, the Hamiltonian of system do not commutes with all the generators ($g_i$) of the symmetry group (G), $[\hat{H}, g_i] \neq 0$, and none of the states have good symmetry. In-between these limiting cases, intermediate symmetry structures which called partial (dynamical) symmetries, for which the symmetry is neither exact nor completely broken. The partial dynamical symmetry (PDS) has been introduced [6-11] in order to a further enlargement of the fundamental concepts of the exact and dynamical symmetries. PDS provides an intermediate symmetry structure which corresponds to a particular symmetry breaking, but preserves the useful aspects of a dynamical symmetry for a part of the system. The advantage of using interactions with a PDS is that they can be introduced, in a controlled manner, without destroying results previously obtained with a DS for a segment of the spectrum [9]. One important aspect of PDS is their ability to serve as a practical tool for calculation of observables in real systems. The mathematical aspects and algorithm for constructing the Hamiltonians with partial dynamical symmetry has been developed in Ref.[6] and further elaborated to Hamiltonians with higher-order terms presented in Refs.[10,11] by Leviatan *et al*. Also, the relevance of $SU(3)$-PDS to the spectroscopy of $^{168}Er$ and $^{160}Dy$ nuclei have been described in Refs.[7,12] and showed that, this nuclei can be a good examples of $SU(3)$-PDS, i.e. the resulting PDS calculations are found to be in excellent agreement with experimental data. The aim of present work is to tests of nuclear partial dynamical $SU(3)$ symmetry for axially deformed prolate nuclei in the rare earth and actinide regions, in order to illustrate the usefulness of $SU(3)$-PDS for predicting properties of these nuclei. Here we have carried out extensive study of $SU(3)$-PDS, studying 51 even-even nuclei, to see if the PDS is broadly applicable, and where, how, and in which nuclei it breaks down.

This article is organized as follows. In Sec. II the theoretical framework of $SU(3)$-PDS calculation for axially deformed prolate nuclei is briefly described. In Sec. III we present numerical results and detailed discussion for evaluating the best fitted nuclei in PDS and DS scheme, including some significant parameters. A brief summary is given in Sec. IV.

## II. Model

The $SU(3)$-DS is an appropriate symmetry structure introduced in the interacting boson model (IBM) framework for describing the axially deformed nuclei which based on the pioneering works of Elliott [13]. The IBM [14-16] provides a rich algebraic structure to illustrate the implications of the partial dynamical symmetry which is widely used in description of low-lying collective states in nuclei. Therefore, we consider the relevant aspects of this model which related to the $SU(3)$-PDS. The IBM description of an axially deformed nucleus is the $SU(3)$ limit which describes a symmetric rotor with degenerate $\beta$ and $\gamma$ bands. The basis states in this limit are labeled by $\left|[N](\lambda,\mu)KLM\right\rangle$ where $N$ is the total number of bosons, $(\lambda,\mu)$ denote the $SU(3)$ irreducible representations (irreps), $L$ is the angular momentum and $K$ is the multiplicity label. This extra quantum number, i.e. $K$, which corresponds geometrically to the projection of the angular momentum on the symmetry axis, is necessary for complete classification. Each $K$-value in a given $SU(3)$ irrep $(\lambda,\mu)$, is associated with a rotational band and in different $K$-bands, states with the same $L$ are degenerate. The ground $g(K=0)$ band of an axially deformed prolate nucleus which described by irrep $(2N,0)$, is the lowest $SU(3)$ irrep. On the other hand, both the $\beta(K=0^+_2)$ and $\gamma(K=2^+_1)$ bands, which is used to describe the lowest excited bands, span the irrep $(2N-4,2)$ and therefore, the states in $\beta$ and $\gamma$ bands with the same $L$ are degenerate. This undesired $\beta-\gamma$ degeneracy, which is a characteristic feature of the $SU(3)$ limit in IBM framework, can be lifted by adding an extra term from other chains to the $SU(3)$ Hamiltonian, although this kind of $K$-band degeneracy is not commonly observed in deformed nuclei [9]. In the empirical spectra of most deformed nuclei the $\beta$ and $\gamma$ bands are not degenerate and thus, to conform the experimental data, one is compelled to break $SU(3)$ symmetry. Such a $SU(3)$ symmetry breaking introduced by Warner, Casten and Davidson (WCD) [15] or similar approach was taken in the consistent $Q$ formalism (CQF) by the same authors [16] in order to lift the undesired $\beta-\gamma$ degeneracy. In these procedures, where an additional term from other chains was added to the $SU(3)$ Hamiltonian, the $SU(3)$ symmetry is completely broken, all eigenstates are mixed and none of virtues are retained. In contrast, Leviatan [7], have introduced the partial dynamical $SU(3)$ symmetry in which corresponds to a particular $SU(3)$ symmetry breaking, but preserves the useful aspects of a dynamical symmetry, e.g., the solvability for a part of the system. Hamiltonian of $SU(3)$-DS composed of a linear combination of the Casimir operators of $SU(3)$ and $O(3)$ groups. A two-body $SU(3)$-PDS Hamiltonian in the framework of IBM has the form [9]

$$\hat{H}_{PDS} = \hat{H}(h_0,h_2) + C\hat{C}_{O(3)} = h_0 P_0^\dagger P_0 + h_2 P_2^\dagger \cdot \tilde{P}_2 + C\hat{C}_{O(3)} \qquad , \qquad (1)$$

Where $\hat{H}(h_0,h_2)$ is a two-body Hamiltonian with partial $SU(3)$ symmetry, $P_0^\dagger = d^\dagger \cdot d^\dagger - 2(s^\dagger)^2$ and $P_{2\mu}^\dagger = 2d_\mu^\dagger s^\dagger + \sqrt{7}(d^\dagger d^\dagger)_\mu^{(2)}$ are the boson-pair operators in IBM with angular momentum $L=0$ and $2$, respectively and $\hat{C}_{O(3)}$ denotes the Casimir operator of $O(3)$ group. For $h_0 = h_2$ case, the $\hat{H}(h_0,h_2)$ forms a $SU(3)$ scalar related to the Casimir operator of $SU(3)$ while for $h_0 = -5h_2$, it is an $SU(3)$ tensor, namely $(\lambda,\mu)=(2,2)$. Although the $\hat{H}(h_0,h_2)$ is not a $SU(3)$ scalar, it has a subset of solvable states with good $SU(3)$ symmetry. The additional $O(3)$ rotational term which converts the partial $SU(3)$ symmetry into $SU(3)$-PDS, contributes an $L(L+1)$ splitting and has no effect on wave functions and consequently, the undesired $\beta-\gamma$ degeneracy can be lifted. According to the prescription which introduced in Ref.[9], the solvable states of $\hat{H}_{PDS}$ which preserve the $SU(3)$ symmetry, are members of the ground $g(K=0)$ and $\gamma^k(K=2k)$ bands and have the form

For $g(K=0):\left|N,(2N,0),K=0,L\right\rangle \quad\Rightarrow E_{PDS}=CL(L+1),L=0,2,...,2N$ , (2a)

For $\gamma^k(K=2k):\qquad \left|N,(2N-4k,2k),K=2k,L\right\rangle \qquad\Rightarrow$

$E_{PDS}=6h_2k(2N-2k+1)+CL(L+1),\qquad L=K,K+1,...,(2N-2k)$ , (2b)

$\hat{H}_{PDS}$, i.e. Eq. (1), is specified by three parameters, namely $C,h_2$ and $h_0$, which the values of $C$ and $h_2$ were extracted from the experimental energy differences $[E(2_g^+)-E(0_g^+)]$ and $[E(2_\gamma^+)-E(2_g^+)]$ respectively [9]. In PDS calculation, the parameter $h_0$ was varied in order to reproduce the bandhead energy of the $\beta$ band. These parameters are given for each nucleus in Table 1.

## III. RESULTS AND DISCUSSION

In this study we would like to address the problem of identifying nuclei exhibiting the axially deformed prolate structure by using $SU(3)$-PDS. For this purpose in this section, we will follow the systematic analysis outlined by Kern *et al.* [17] and Abul-magd *et al* [18], which were used systematic IBM calculations to identify nuclei that have a better fit with $U(5)$-DS and $SU(3)$-DS, respectively. Here we carry out a similar study on nuclei which assumed to satisfy the criteria of the $SU(3)$-PDS. The present study involves data on low-lying levels of selected the axially deformed prolate even–even nuclei which are taken from the Nuclear Data Sheets [19] until August 2012. In This analysis, we have considered the nuclei in which the ground state band, first excited $K = 0^+$ and $K = 2^+$ ($\beta$ and $\gamma$) bands are definite and also, the spin-parity $J^\Pi$ assignments of consecutive levels of these bands are definite. According to the latest available experimental data in a few cases where the $\beta$- and $\gamma$- bands are not assigned, their analysis have not been considered here. In this manner, we have found 51 nuclei with definite $R_{4/2} \geq 3.00$ ratio which have given in Table 1.

Generally, the overall structure of a typical rotational even-even nucleus can be interpreted by taking into account a few significant observables. In terms of the energies, emphasis was placed on the $R_{4/2}$ ratio which is a fundamental observable to describe the structure of nucleus. In this analysis, we have not restrict to select nucleus with $R_{4/2} = 3.33$, which is indicate an ideal rotational nucleus in $SU(3)$-DS limit of IBM. As we know that most deformed rotational nuclei do not meet this fixed value. Instead, we examine the range $R_{4/2} \geq 3.00$ which, indeed, marks a value typical of the onset of rotational structure, in order to have a sufficient number of nuclei for extensive analysis of $SU(3)$-PDS and to see if the PDS is broadly applicable, and where, how, and in which nuclei it breaks down.

Another significant sensitive quantity is the $P$-factor [20-23], which is provides a general and physically meaningful explanation for the development of the collectivity and deformation in the structure of nuclei, in which a measure of the average number of interactions of each valence nucleon with those of the other type is given. This important parameter, with regardless of the mass regions, indicate the onset of deformation, i.e. $R_{4/2} \approx 3.00$, when reaches $\approx 4 - 5$, which means that a deformed nuclei with fewer than four valance nucleons of either type can never become deformed. We need at least four or five valance p-n interaction to overcome one pairing interaction, since the pairing interaction strength is about $1 Mev$, and each p-n interaction has strength $\approx 200\text{-}250 \, kev$. The main criteria which has used in this survey to searching for deformed rotational nuclei is based on the $R_{4/2} \geq 3.00$ and $P > 4$, which indicates onset of deformation and rotational structure in nuclei. The $P$-factor for considered nuclei is listed in Table 1.

One of the important features of the $SU(3)$ dynamical symmetry is the degeneracy of levels having the same spin of the $\beta$- and $\gamma$-bands of the lowest excitation energy $K = 0^+$, and $K = 2^+$ of irreducible representations $(\lambda, \mu) = (2N - 4, 2)$, respectively [2]. As have explained extensively in Refs.[7-11], the experimental spectrum of deformed nuclei and especially, the $\beta \left( K = 0_2^+ \right)$ and $\gamma \left( K = 2_1^+ \right)$ bands are not degenerate. On the other hand, one can expect, the spectrum of an exact $SU(3)$-DS which is obtained by $h_0 = h_2$ and therefore indicate degeneracy of these bands deviates considerably from the empirical data. The lifted $\beta$-$\gamma$ degeneracy governed by the predictions of $SU(3)$-PDS, show an improvement over the schematic description of exact $SU(3)$ dynamical symmetry. In order to indicate $\beta$-$\gamma$ degeneracy in considered nuclei, we use the empirical systematics of the energy difference between $2^+_\beta$ and $2^+_\gamma$ which the $SU(3)$-DS value for this difference is zero [24]. According to the Ref. [18], we use a degeneracy parameter, i.e. $\delta$, in order to indicate $\beta$-$\gamma$ degeneracy in considered nuclei which can be calculate as

$$\delta = |E(2^+_\beta) - E(2^+_\gamma)| \leq 100 \ keV \qquad , \qquad (3)$$

Which means that $2^+_\beta$ and $2^+_\gamma$ states could be considered degenerate when the energy difference between these states is around $100 \ kev$. We have reported this quantity for considered nuclei in Table 1 which the seventeen of these nuclei meet this characteristic property.

The quality of the fit is summarized by the root-mean-square (RMS) deviation, absolute average deviation, $\Delta$, and quality factor, $Q$ [17-18], between the experimental and calculated level energies. These quantities value obtained for two different approaches, e.g. $SU(3)$-PDS vs. $SU(3)$-DS, which are given in Table 1. The quality factor which are defined by $Q = (N_L - b)^{-1} \sum_i W_i (E_i^{exp} - E_i^{fit})^2$ where $b$ is the number of the adjustable parameters and $W_i = 0.01$ is a weighing factor chosen to correspond to a uniform uncertainty of $10 \ kev$ on the level energies. In this analysis, all energy levels of each nucleus from ground, $\beta$ and $\gamma$ bands which is appearing below the $2 \ Mev$, have been included. The number of the available experimental energy levels, i.e. $N_L$ which was reported in Table 1, was variable ranging between 9 in e.g. $^{228}Ra$ to 24 in $^{232}Th$. We have considered the energy levels of a nucleus to satisfy the $SU(3)$-$PDS$ formula if $\Delta \leq 100 \ kev$ and $Q \leq 150 \ kev$ which was considered by Ref.[18] for $SU(3)$- DS to represent a best fitted nuclei and corresponds to deviation of less than 5% and better. In this view, with few exceptions, e.g. transitional nuclei, for all levels almost near 2 $Mev$, all nuclei which have considered are well fitted in PDS scheme. The overall agreement is excellent, with the PDS predictions always agreeing with the experimental to within less than 5% or better. Only in the following nuclei from transitional region $^{152}Sm$, $^{154}Gd$, $^{160}Er$, $^{168}Hf$, $^{178}Os$, $^{180}Os$, the deviation from experimental counterparts is near 10%. However, with these quality parameters only less than half of these nuclei have a good description in DS which is due to degeneracy of levels. The $Q$, $\Delta$ and RMS quantities are sensitive to the number of levels involved in fitting procedure, for this reason, we restrict $N_L$ to be equal from each bands to make prediction of models more comparable and have an extensive analysis.

In all isotopic chains in rare earth region with $62 \leq Z \leq 76$, we have found that the PDS (RMS) deviation value is decreased as neutron number is increased, with few exceptions $^{186}_{76}Os$, $^{184}_{74}W$, $^{186}_{74}W$, which is due to the passing a deformed subshell gap at $N$=108 [20-23] in which a change in tendency occurs and (RMS) deviation value increase. A similar trend is observed for all isotonic chains in this region with $90 \leq N \leq 112$, with an exception, $^{154}_{64}Gd$, which is due to the obliteration of the $Z$=64 proton subshell closure at about $N$=90 [24], in which the RMS value of PDS is decreased as proton number is decreased. (Fig. 1b)

As $N$ increased from $N$=140 to 152 in isotopic chains in actinide region with $88 \leq Z \leq 98$, we have found a same trends like rare earth region in which the PDS (RMS) deviation value decreased around $N$=144 and then a slight growing up after $N$=144 and decreased in $N$=152, which may be reflect the appearing and disappearing effects of some subshell gaps like $N$=142, $N$=144 subshell gaps and the well-recognized $N$=152 gap [25-31] which could influence the fine structure in this region. For isotonic chains in this region with $140 \leq N \leq 152$, we have observed that the RMS values of PDS have decreased as proton number is increased in $N$=140-150 chains and start to increase in $N$=152 chain as $Z$ increased (Fig. 1d). As indicated in Figs. 1(a) to 1(d) for all isotopic and isotonic chains in rare earth and actinide regions, despite the regular pattern in PDS predictions, DS has an irregular contour which is due to the $\beta - \gamma$ degeneracy prediction in DS scheme for all of these nuclei.

As shown in Figs. 2(a) and 2(b) for rare earth and Figs. 3(a) and 3(b) for actinide regions, an interesting implication from this analysis is that in all isotopic and isotonic chains we have found an exciting behavior by raising quadrupole deformations, with few exceptions as mentioned above, which lead to a precise prediction in PDS scheme., e.g. PDS has a most accurate prediction in well-deformed nuclei in each isotope chains. A maximum value of deformation in each isotopic and isotonic chains leads to minimum value of RMS deviation in PDS scheme. For the nuclei which included in this survey, all chains begin as transitional with $R_{4/2}$ around 3.00 and moves towards rotational ($R_{4/2} \rightarrow 3.33$) in which the RMS value of PDS decreased as the $R_{4/2}$ value is increased and the PDS has a most accurate prediction near the $R_{4/2} \approx 3.33$ as indicated in Figs. 2(c) and 2(d) for rare earth and Figs. 3(c) and 3(d) for actinide regions. In all Isotopic chains in rare earth and actinide regions, we have found that the RMS values of PDS have decreased, with exceptions in some cases as mentioned above due to appearing or disappearing subshell gaps, as the $P$-factor value is increased. It is equally evident that the exact inverse behavior occurs for all isotonic chains as indicated in Figs. 2(e) and 2(f) for rare earth and Figs. 3(e) and 3(f) for actinide regions. Another interesting point concerns the $\beta - \gamma$ degeneracy which in nuclei with this $\beta - \gamma$ property, PDS has a similar result like DS. As an evident from Fig. 4 in all nuclei which have prescribed in this survey one can see a similar prediction from degenerate nuclei which have $\delta \leq 100 kev$, as $\delta$ value increased the RMS deviation of DS grow up rapidly and have linear trend by increasing $\delta$ value in while PDS has not sensitive to these parameter and worked well in all cases.

We have 30 radioactive nuclei in this analysis which is 15 of them with $\alpha$ decay, 4 nuclei with $\beta^-$, 2 with $2\beta^-$ decay and also we have 9 nuclei with $\varepsilon$ decay mode which is indicate the combination of electron capture and $\beta^+$. In $\alpha$ decay of radioactive nuclei, we have a fluctuation in deviation of PDS due to the appearing some subshell gap effects in both rare earth and actinide regions as well as insufficient nuclei in each chains, have been concerned the study of half-lives in these nuclei in PDS scheme. Also we have an insufficient cases in $\beta^-$ and double-beta decay to deduce a perfect result in such important radioactive cases. As apparent from Figs. 5(c) and 5(d), in all radioactive nuclei with $\varepsilon$ decay we have observed that in all isotopic and isotonic chains, the RMS value in PDS increased by growing up the half-life of these nuclei which is due to the increasing the $\beta_2$ -deformation parameter and we have a minimum deviation in $^{172}Hf$ which has a maximum half-life value.

In Table 1, we have reported some important parameters like the $\beta_2$ -deformations, $P$-factor, and dominant decay modes with $T_{1/2}$ to make more predictive analysis for considered nuclei. The $\beta_2$ values are taken from Ref. [32] and decay modes with $T_{1/2}$ from Ref. [33].

Table 1. Parameters of PDS and comparisons of the quality factor, $Q$, the absolute average deviation, $\Delta$ and the root mean-square (RMS) deviation for fitted energy levels, $N_L$, in two different approaches($SU(3)$-PDS vs. $SU(3)$-DS) and some important parameters like the $P$-factor, $T_{1/2}$, decay modes and $\delta$ are shown for nuclei belonging to the interval $R_{4/2} \geq 3.00$.

| nucleus | $N_B$ | $R_{4/2}$ | $p$ | $N_L$ | $T_{1/2}$ | Decay Mode | $h_0$ (keV) | $h_2$ (keV) | C (keV) | $\Delta$(keV) PDS | $\Delta$(keV) DS | RMS(keV) PDS | RMS(keV) DS | Q(keV) PDS | Q(keV) DS | $\beta_2$ | $\delta$ (keV) |
|---|---|---|---|---|---|---|---|---|---|---|---|---|---|---|---|---|---|
| $^{152}_{62}Sm$ | 10 | 3.009 304 | 4.8 | 17 | STABLE | - | 4.33 | 8.45 | 20.29 | 129.82 | 226.45 | 222.69 | 334.00 | 492.93 | 1113.6 | 0.243 | 275 |
| $^{154}_{62}Sm$ | 11 | 3.254 6 | 5.46 | 17 | STABLE | - | 7.10 | 10.77 | 13.66 | 53.56 | 134.43 | 95.13 | 201.60 | 87.50 | 404.46 | 0.270 | 263 |
| $^{154}_{64}Gd$ | 11 | 3.014 520 | 5.09 | 17 | STABLE | - | 4.26 | 6.93 | 20.51 | 128.88 | 196.35 | 213.50 | 280.12 | 452.84 | 782.72 | 0.243 | 180 |
| $^{156}_{64}Gd$ | 12 | 3.239 15 | 5.83 | 17 | STABLE | - | 7.07 | 7.71 | 14.82 | 54.50 | 72.74 | 95.33 | 116.4 | 87.88 | 133.6 | 0.271 | 24 |
| $^{158}_{64}Gd$ | 13 | 3.288 24 | 6.46 | 16 | STABLE | - | 7.80 | 7.38 | 13.25 | 40.41 | 38.44 | 68.10 | 63.97 | 43.37 | 38.92 | 0.271 | 72 |
| $^{158}_{66}Dy$ | 13 | 3.206 08 | 6.15 | 17 | STABLE | - | 6.69 | 5.64 | 16.48 | 59.68 | 67.34 | 97.79 | 100.30 | 92.63 | 98.60 | 0.262 | 139 |
| $^{160}_{64}Gd$ | 14 | 3.302 2 | 7.00 | 17 | > 3.1E + 19 y | $2\beta^-$ | 9.01 | 5.63 | 12.54 | 35.66 | 95.08 | 54.26 | 159.52 | 26.44 | 252.49 | 0.280 | 389 |
| $^{160}_{66}Dy$ | 14 | 3.270 | 6.86 | 17 | STABLE | - | 8.71 | 5.42 | 14.46 | 54.22 | 111.65 | 94.51 | 173.39 | 86.33 | 298.65 | 0.272 | 383 |
| $^{160}_{68}Er$ | 12 | 3.099 | 5.83 | 12 | 28.58 h | $\varepsilon$ | 6.66 | 5.27 | 20.96 | 99.08 | 119.39 | 186.47 | 193.60 | 344.73 | 372.81 | 0.253 | 153 |
| $^{162}_{64}Gd$ | 15 | 3.301 | 7.47 | 12 | 8.4 m | $\beta^-$ | 9.61 | 4.55 | 11.93 | 24.37 | 124.48 | 47.95 | 251.18 | 19.99 | 628.95 | 0.291 | 628 |
| $^{162}_{66}Dy$ | 15 | 3.293 | 7.47 | 18 | STABLE | - | 11.56 | 4.64 | 13.44 | 46.83 | 177.95 | 71.84 | 337.76 | 48.61 | 1138.8 | 0.281 | 840 |
| $^{162}_{68}Er$ | 13 | 3.230 | 6.46 | 17 | STABLE | - | 7.79 | 5.32 | 17.00 | 54.39 | 90.66 | 88.31 | 130.78 | 74.99 | 169.05 | 0.272 | 270 |
| $^{164}_{68}Er$ | 14 | 3.276 | 7.00 | 18 | STABLE | - | 8.75 | 4.74 | 15.23 | 55.59 | 123.86 | 81.67 | 193.14 | 63.70 | 371.05 | 0.273 | 454 |
| $^{166}_{66}Dy$ | 17 | 3.310 33 | 8.48 | 13 | 81.6 h | $\beta^-$ | 6.47 | 3.94 | 12.76 | 21.66 | 71.99 | 40.16 | 137.72 | 13.13 | 187.67 | 0.293 | 351 |
| $^{166}_{68}Er$ | 15 | 3.288 | 7.47 | 21 | STABLE | - | 10.14 | 4.05 | 13.42 | 74.54 | 208.33 | 109.74 | 331.29 | 117.44 | 1095.5 | 0.283 | 742 |
| $^{166}_{70}Yb$ | 13 | 3.228 | 6.46 | 17 | 56.7 h | $\varepsilon$ | 7.26 | 5.53 | 17.06 | 63.10 | 82.53 | 105.4 | 120.8 | 108.1 | 144.0 | 0.274 | 211 |
| $^{168}_{70}Yb$ | 14 | 3.266 | 6.85 | 16 | STABLE | - | 7.54 | 5.53 | 14.62 | 58.06 | 91.97 | 105.64 | 137.16 | 108.60 | 186.14 | 0.284 | 250 |
| $^{168}_{68}Er$ | 16 | 3.309 | 7.88 | 18 | STABLE | - | 7.51 | 3.98 | 13.30 | 38.40 | 110.84 | 59.12 | 190.35 | 31.96 | 360.33 | 0.294 | 455 |
| $^{168}_{72}Hf$ | 12 | 3.109 8 | 5.83 | 14 | 25.95 m | $\varepsilon$ | 7.08 | 5.44 | 20.68 | 82.23 | 104.76 | 167.07 | 177.37 | 276.13 | 312.62 | 0.254 | 182 |
| $^{170}_{68}Er$ | 17 | 3.309 | 8.24 | 20 | STABLE | - | 4.39 | 4.32 | 13.09 | 34.54 | 33.11 | 56.06 | 53.61 | 28.43 | 26.74 | 0.296 | 26 |
| $^{170}_{70}Yb$ | 15 | 3.292 | 7.20 | 16 | STABLE | - | 5.86 | 6.10 | 14.04 | 60.31 | 67.87 | 96.80 | 105.17 | 90.71 | 108.62 | 0.295 | 7 |
| $^{170}_{72}Hf$ | 13 | 3.194 | 6.15 | 12 | 16.01 h | $\varepsilon$ | 5.58 | 5.73 | 16.8 | 58.99 | 61.72 | 113.46 | 114.30 | 125.73 | 128.66 | 0.274 | 25 |
| $^{172}_{70}Yb$ | 16 | 3.305 | 7.50 | 16 | STABLE | - | 4.70 | 7.97 | 13.12 | 30.49 | 153.01 | 54.70 | 247.97 | 26.93 | 612.93 | 0.296 | 348 |
| $^{172}_{72}Hf$ | 14 | 3.2476 | 6.42 | 14 | 1.87 y | $\varepsilon$ | 4.75 | 6.04 | 15.87 | 35.61 | 66.70 | 69.44 | 104.16 | 45.22 | 106.50 | 0.284 | 122 |
| $^{174}_{72}Hf$ | 15 | 3.2685 | 7.47 | 15 | 2.0E+15 y | $\alpha$ | 3.63 | 6.52 | 15.16 | 62.02 | 177.62 | 103.15 | 274.74 | 103.40 | 752.83 | 0.285 | 326 |
| $^{176}_{72}Hf$ | 16 | 3.2845 | 7.00 | 16 | STABLE | - | 5.61 | 6.73 | 14.72 | 48.50 | 93.32 | 88.28 | 151.28 | 74.94 | 226.87 | 0.277 | 115 |
| $^{176}_{74}W$ | 14 | 3.215 | 5.71 | 14 | 2.5 h | $\varepsilon$ | 4.65 | 5.75 | 18.05 | 70.44 | 105.6 | 118.5 | 154.6 | 137.4 | 237.0 | 0.266 | 110 |
| $^{178}_{72}Hf$ | 15 | 3.29059 | 6.46 | 16 | STABLE | - | 6.88 | 6.21 | 15.53 | 34.00 | 34.64 | 54.30 | 50.97 | 26.48 | 23.97 | 0.278 | 102 |
| $^{178}_{70}Yb$ | 15 | 3.310 | 7.20 | 10 | 74 m | $\beta^-$ | 7.69 | 6.53 | 14 | 14.31 | 46.34 | 23.33 | 70.64 | 2.44 | 47.91 | 0.279 | 183 |
| $^{178}_{74}W$ | 15 | 3.236 | 5.86 | 13 | 21.6 d | $\varepsilon$ | 7.90 | 5.77 | 17.65 | 46.99 | 98.27 | 87.90 | 146.63 | 74.27 | 213.00 | 0.267 | 307 |
| $^{178}_{76}Os$ | 13 | 3.017 | 4.61 | 14 | 5.0 m | $\varepsilon$ | 3.81 | 4.88 | 22.03 | 121.3 | 153.0 | 235.0 | 250.7 | 549.3 | 628.6 | 0.247 | 93 |
| $^{180}_{72}Hf$ | 14 | 3.30652 | 6.43 | 13 | STABLE | - | 6.42 | 6.82 | 15.55 | 16.39 | 26.86 | 28.15 | 40.69 | 4.92 | 14.557 | 0.279 | 16 |
| $^{180}_{76}Os$ | 14 | 3.093 | 4.71 | 16 | 21.5 m | $\varepsilon$ | 4.29 | 4.55 | 22.01 | 119.2 | 119.2 | 213.1 | 209.1 | 451.1 | 435.3 | 0.238 | 39 |
| $^{182}_{74}W$ | 13 | 3.290782 | 5.54 | 14 | STABLE | - | 7.18 | 7.47 | 16.68 | 24.49 | 23.59 | 41.40 | 40.51 | 14.14 | 14.41 | 0.259 | 36 |
| $^{184}_{74}W$ | 12 | 3.27350 | 5.33 | 15 | STABLE | - | 7.57 | 5.73 | 18.53 | 46.91 | 80.39 | 78.64 | 110.71 | 58.84 | 120.57 | 0.240 | 218 |
| $^{184}_{76}Os$ | 12 | 3.203 | 4.5 | 15 | > 5.6E13 y | $\alpha$ | 7.87 | 5.96 | 19.96 | 49.84 | 98.28 | 98.05 | 144.85 | 93.15 | 207.83 | 0.229 | 261 |
| $^{186}_{74}W$ | 11 | 3.233 | 5.09 | 14 | > 2.3E+19 y | $2\beta^-$ | 7.59 | 4.88 | 20.43 | 37.07 | 88.28 | 63.89 | 131.76 | 37.82 | 171.63 | 0.230 | 292 |
| $^{186}_{76}Os$ | 11 | 3.16485 | 4.36 | 15 | 2.0E+15 y | $\alpha$ | 9.55 | 5.00 | 22.85 | 62.49 | 133.90 | 97.84 | 196.80 | 92.73 | 385.33 | 0.220 | 440 |
| $^{228}_{88}Ra$ | 11 | 3.207 | 4.2 | 9 | 5.75 y | $\beta^-$ | 5.09 | 6.20 | 10.63 | 13.57 | 46.37 | 22.36 | 72.19 | 2.00 | 50.12 | 0.180 | 75 |
| $^{230}_{90}Th$ | 11 | 3.2711 | 5.09 | 22 | 7.54E+4 y | $\alpha$ | 4.32 | 5.77 | 8.87 | 73.44 | 102.5 | 107.1 | 136.1 | 111.7 | 183.4 | 0.198 | 104 |
| $^{232}_{90}Th$ | 12 | 3.2838 | 5.33 | 24 | 1.40E10 y | $\alpha$ | 4.94 | 5.33 | 8.22 | 68.45 | 77.77 | 101.2 | 108.0 | 99.43 | 114.7 | 0.207 | 11 |
| $^{232}_{92}U$ | 12 | 3.2911 | 5.39 | 18 | 68.9 y | $\alpha$ | 4.23 | 5.936 | 7.92 | 55.92 | 110.39 | 103.85 | 156.57 | 104.85 | 243.14 | 0.207 | 132 |
| $^{234}_{92}U$ | 13 | 3.29559 | 6.15 | 21 | 2.455E+5 y | $\alpha$ | 4.84 | 5.88 | 7.24 | 39.49 | 65.35 | 79.80 | 99.32 | 60.69 | 96.65 | 0.215 | 75 |
| $^{236}_{92}U$ | 14 | 3.30380 | 6.42 | 16 | 2.342E7 y | $\alpha$ | 5.38 | 5.63 | 7.54 | 38.47 | 43.59 | 78.59 | 79.97 | 58.77 | 61.96 | 0.215 | 2 |
| $^{238}_{92}U$ | 15 | 3.3035 | 6.67 | 22 | 4.468E9 y | $\alpha$ | 4.80 | 5.83 | 7.486 | 45.75 | 62.55 | 77.56 | 91.55 | 57.16 | 81.82 | 0.215 | 94 |
| $^{238}_{94}Pu$ | 15 | 3.3114 | 7.20 | 15 | 87.7 y | $\alpha$ | 5.01 | 5.65 | 7.346 | 30.41 | 41.41 | 60.35 | 67.19 | 33.42 | 43.15 | 0.215 | 45 |
| $^{240}_{94}Pu$ | 16 | 3.3087 | 7.50 | 19 | 6561 y | $\alpha$ | 3.783 | 5.88 | 7.13 | 41.03 | 140.04 | 69.76 | 201.32 | 45.66 | 403.33 | 0.223 | 237 |
| $^{242}_{94}Pu$ | 17 | 3.307 | 7.77 | 12 | 3.75E+5 y | $\alpha$ | 4.35 | 5.34 | 7.42 | 39.35 | 61.60 | 73.25 | 92.27 | 50.66 | 83.15 | 0.224 | 110 |
| $^{246}_{96}Cm$ | 19 | 3.3135 | 8.84 | 14 | 4706 y | $\alpha$ | 5.29 | 4.87 | 7.14 | 39.36 | 47.51 | 69.57 | 73.22 | 45.40 | 51.62 | 0.234 | 86 |
| $^{248}_{96}Cm$ | 20 | 3.309 | 9.10 | 14 | 3.48E+5 y | $\alpha$ | 4.62 | 4.29 | 7.23 | 31.73 | 41.86 | 62.66 | 66.77 | 36.26 | 42.59 | 0.235 | 77 |

| $^{250}_{98}Cf$ | 21 | 3.321 0 | 9.91 | 10 | 13.08 $y$ | $\alpha$ | 4.86 | 4.02 | 7.12 | 3.53 | 30.21 | 5.55 | 61.58 | 2.69 | 35.92 | 0.245 | 158 |

## IV. CONCLUSIONS

The aim of this paper was to illustrate the usefulness of the interacting boson model with $SU(3)$-PDS in its simplest version, the IBM-1, for predicting the properties of axially deformed prolate nuclei in the rare earth and actinide regions.

From these Figures and Tables, one can conclude, the determined results indicate the elegance of the fits presented in this kind of intermediate symmetry structure and they suggest the success of the estimation processes. Since, the Partial dynamical symmetry lifts the remaining degeneracy between $\beta$ and $\gamma$ bands but preserves the symmetry of the selected states, therefore, the acceptable degree of agreement between the predictions of this approach and the experimental counterparts, confirm the relevance of $SU(3)$-PDS to the spectroscopy of considered deformed nuclei.

In summary, we considered the energy levels and the relation of some quantities such as neutron (proton) number, quadruple deformation parameter with the uncertainty measure in the $SU(3)$-PDS framework for 51 deformed nuclei. The validity of the presented parameters, i.e. $h_0, h_2$ and $C$, has been investigated and it is seen that there is an existence of a satisfactory agreement between the presented results and experimental data. We may conclude that the general characteristics of the considered nuclei are well accounted in this study and the idea of the lifted $\beta - \gamma$ degeneracy by $SU(3)$-PDS for these nuclei, is supported. The reduction of RMS values via PDS predictions with increasing the quadrupole deformation and also neutron numbers suggest the ability of this model in describing the structure of deformed nuclei. The obtained results in this study confirm that such particular symmetry breaking is worth extending for investigating the nuclear structure of other nuclei existing in this mass region.

## References

[1]. F. Iachello, A. Arima, The Interacting Boson Mode l. Cambridge University Press, Cambridge (1987).
[2]. F. Iachello, R. D. Levine, Algebraic Theory of Molecules. Oxford University Press, Oxford (1995).
[3]. R. Bijker, F. Iachello, A. Leviatan, Ann. Phys. (N.Y.) 236 (1994) 69.
[4]. F. Iachello, P. Truini, Ann. Phys. (N.Y.) 276 (1999)120.
[5]. K. Kikoin, M.Kiselev, V.Avishai. Dynamical Symmetries for Nanostructures springer-verlage, wien. (2012).
[6].Y. Alhassid, A. Leviatan, J. Phys. A: Math. Gen. 25 (1992) L1265.
[7]. A. Leviatan, Phys. Rev. Lett. **77** (1996) 818.
[8]. A. Leviatan, P. Van Isacker, Phys. Rev. Lett. 89 (2002) 222501.
[9]. A. Leviatan, Prog. Part. Nucl.Phys. 66 (2011)93.
[10]. J.E. Garcia-Ramos, A. Leviatan, P. Van Isacker, Phys. Rev. Lett. 102 (2009) 112502.
[11]. A. Leviatan, J. E. Garcia-Ramos, P. Van Isacker, Phys. Rev. C 87, 021302(R)(2013)

[12]. N. Fouladi, M. A. Jafarizadeh, J. Fouladi, H. Sabri, Cent. Eur. J. Phys. 11,4 (2013) 526-530.

[13]. J.P. Elliott, Proc. Roy. Soc. Lond. Ser.A. 245 (1958) 128 and 562.

[14]. A.Arima, F.Iachello, Ann. Phys.(N.Y.)99(1976)253, Ann. Phys. (N.Y.) 111(1978)201 and Ann. Phys.(N.Y.)123(1979) 468.

[15]. R.F. Casten, D.D. Warner, Rev. Mod. Phys. 60 (1988) 389 and Phys. Rev. C. 28 (1983) 1798.

[16]. D.D. Warner, R.F. Casten, W.F. Davidson, Phys. Rev. C. 24 (1981) 1713.

[17] J. Kern, P.E. Garrett, J. Jolie, H. Lehmann, Nucl. Phys.A 593 (1995) 21.

[18]. A.Y. Abul-Magd, S.A. Mazen, M. Abdel-Mageed, A. Al-Sayed, Nucl.Phys.A. 839(2010)1.

[19] Nucl. Data Sheets, through Vol. 113, Issue 4 (2012).

[20]. R.F. Casten, D.D. Warner, D. S. Brenner, and R.L. Gill, Phys. Rev. Lett. 47 (1981) 1433.

[21] R.F. Casten, Phys. Rev. Lett. 54 (1985) 1991; R.F. Casten, Nucl. Phys. A 443 (1985) 1.

[22] R.F. Casten, D.S. Brenner, P.E. Haustein, Phys. Rev. Lett. 58 (1987) 658.

[23] R.F. Casten, N.V. Zam_r, J. Phys. G 22(1996) 1521.

[24] R.F. Casten, P. von Brentano, A.M.I. Haque, Phys. Rev.C 31 (1985) 1991.

[25] D. Bucurescu and N. V. Zamfir, Phys. Rev.C **87**, 054324 (2013)

[26] S. G. Nilsson, Chin Fu Tsang, A. Sobiczewskim Z. Szymanski, W.Wycek, C. Gustafson, I. L. Lamm, P.M¨oller, and B. Nilsson, Nucl. Phys. A **131**, 1 (1969).

[27] I. Ragnarsson, S. G. Nilsson, and R. K. Sheline, Phys. Rep. **45**,1 (1978).

[28] S. G. Nilsson and I. Ragnarsson, *Shapes and Shells in Nuclear Structure* (Cambridge University Press, Cambridge, UK, 1995).

[29] H. Meldner, Arkiv Physik **36**, 593 (1967).

[30] C. Gustafson, I. L. Lamm, B. Nilsson, and S. G. Nillson, Arkiv Physik **36**, 613 (1967)

[31] A. Ghiorso, S. G. Thompson, G. H. Higgins, B. G. Harvey, and G. T. Seaborg, Phys. Rev. 95, 293 (1954).

[32].P. Moller, J. R. Nix,W. D. Myers, andW. J. Swiatecki, At. Data Nucl. Data Tables 59, 185 (1995).

[33] National Nuclear Data Center,(Brookhaven National laboratory), chart of nuclides. (http://www.nndc.bnl.gov/chart/reColor.jsp?newColor=dm).

## Figure caption

**Figure1 (color online).** Counter plots of PDS and DS, RMS deviation values as a function of neutron and proton number for all isotopic and isotonic chains within (a) and (b) for rare earth region with $62 \leq Z \leq 76$ and $90 \leq N \leq 112$ ,also (c) and(d) actinide region with $88 \leq Z \leq 98$ and $140 \leq N \leq 152$.

**Figure2 (color online).** Counter plots of PDS, RMS deviation values as a function of (a) & (b) $\beta_2$-deformation, (c) &(d) $R_{4/2}$ and (e) &(f) *P*-factor for all isotopic and isotonic chains within rare earth region with $62 \leq Z \leq 76$ and $90 \leq N \leq 112$.

**Figure3(color online).** Counter plots of PDS, RMS deviation value as a function of (a) & (b)$\beta_2$-deformation, (c) &(d) $R_{4/2}$ and (e) &(f)*P*-factor for all isotopic and isotonic chains within actinide region with $88 \leq Z \leq 98$ and $140 \leq N \leq 152$.

**Figure4(color online).** DS and PDS RMS deviation values as a function of degeneracy parameter value, $\delta$.

**Figure5(color online).** Counter plots of DS and PDS, RMS deviation values as a function of neutron and proton number for all isotopic and isotonic chains in radioactive nuclei with $\alpha$ decay (a)&(b) and $\varepsilon$ decay (c)&(d).

Figure 1 (color online).

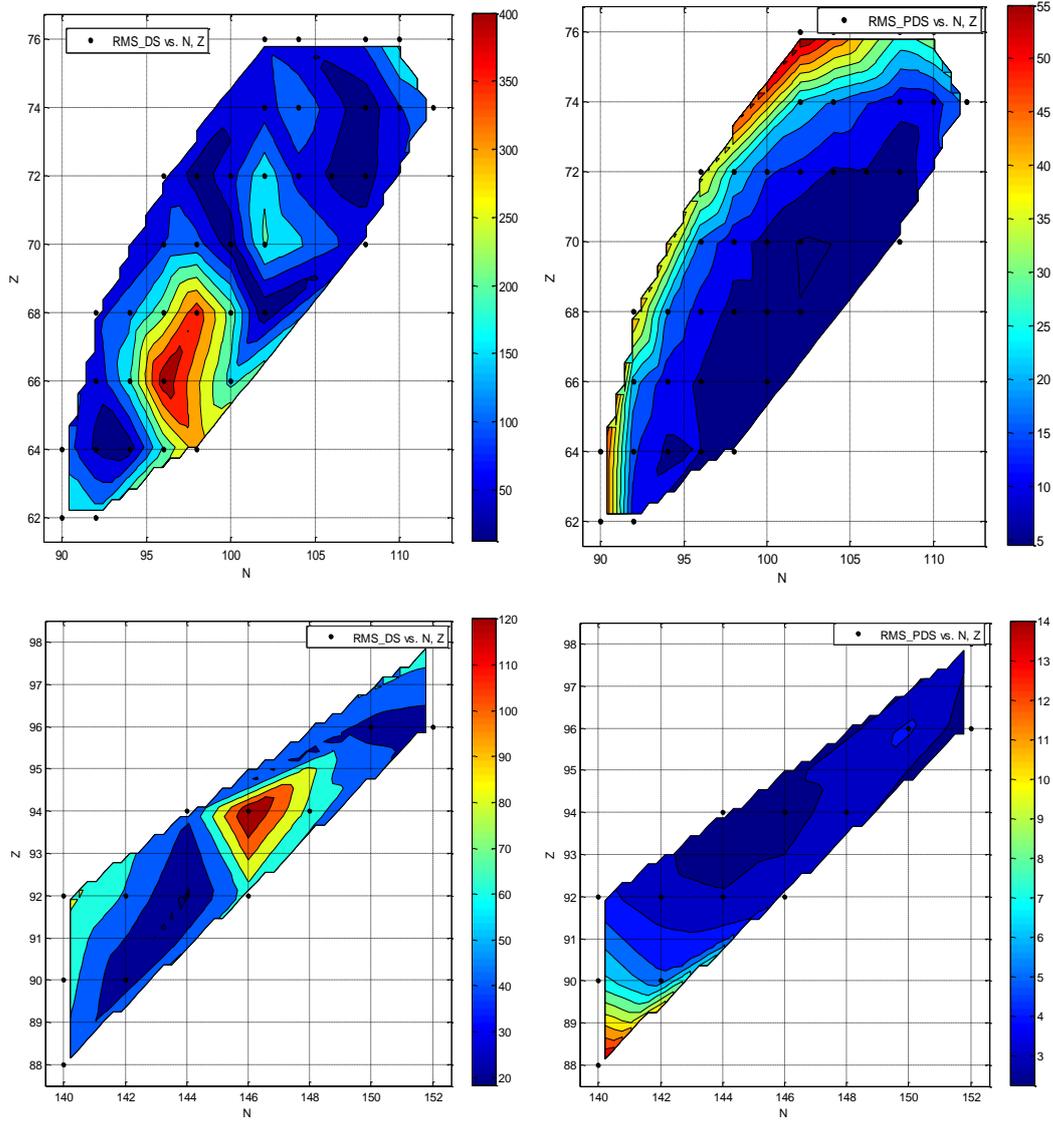

Figure2 (color online).

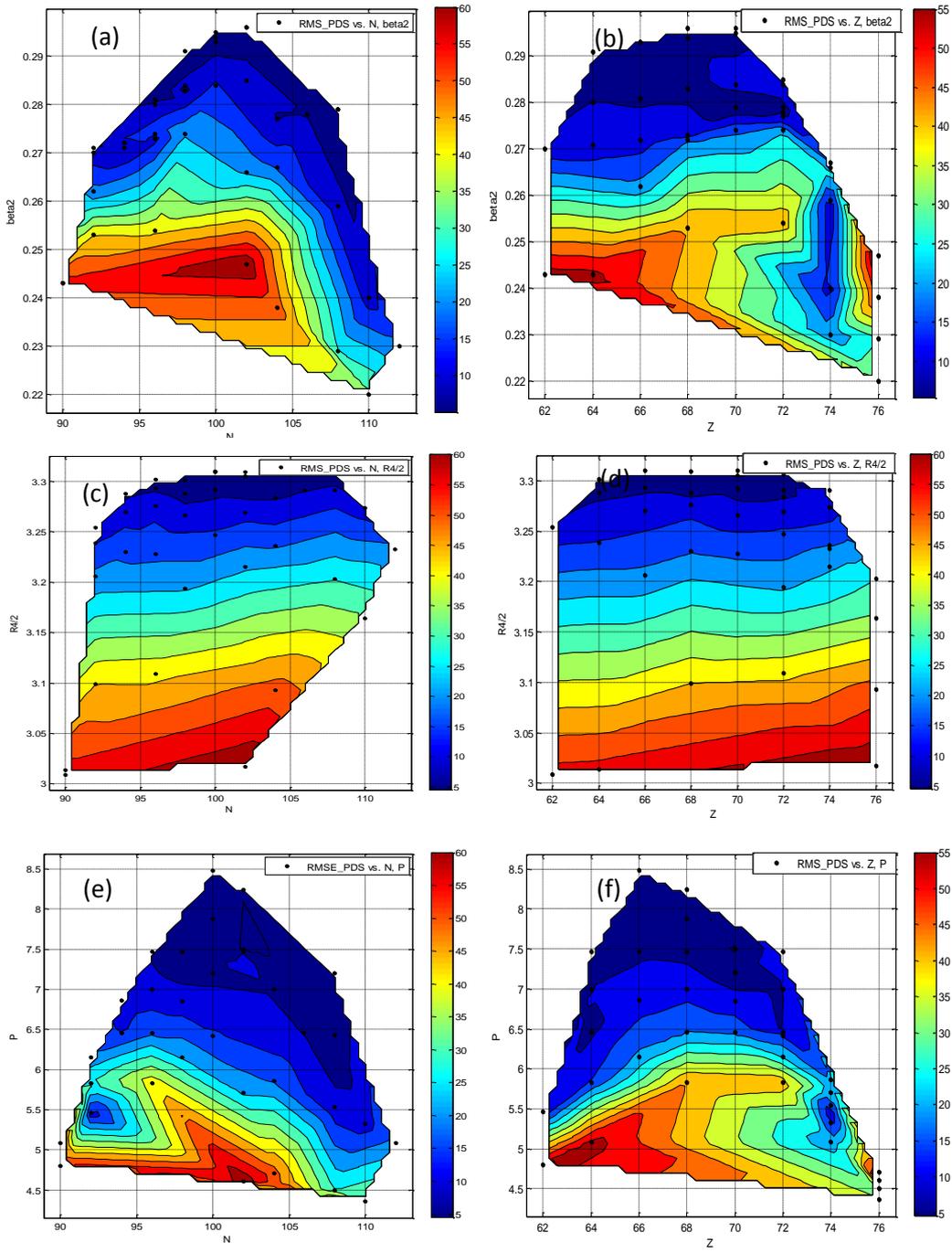

Figure3 (color online).

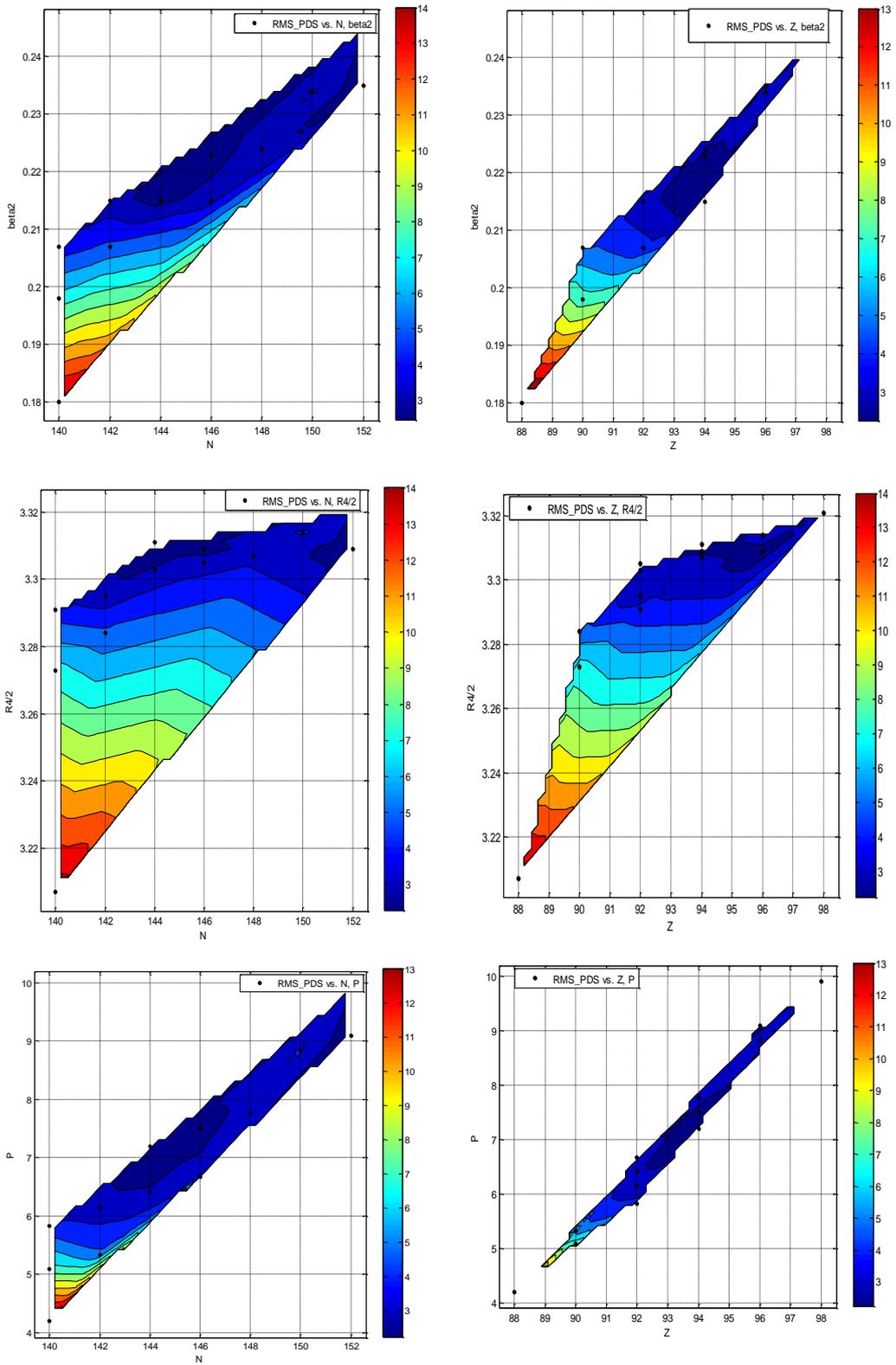

(a)                                    (b)

Figure 4 (color online).

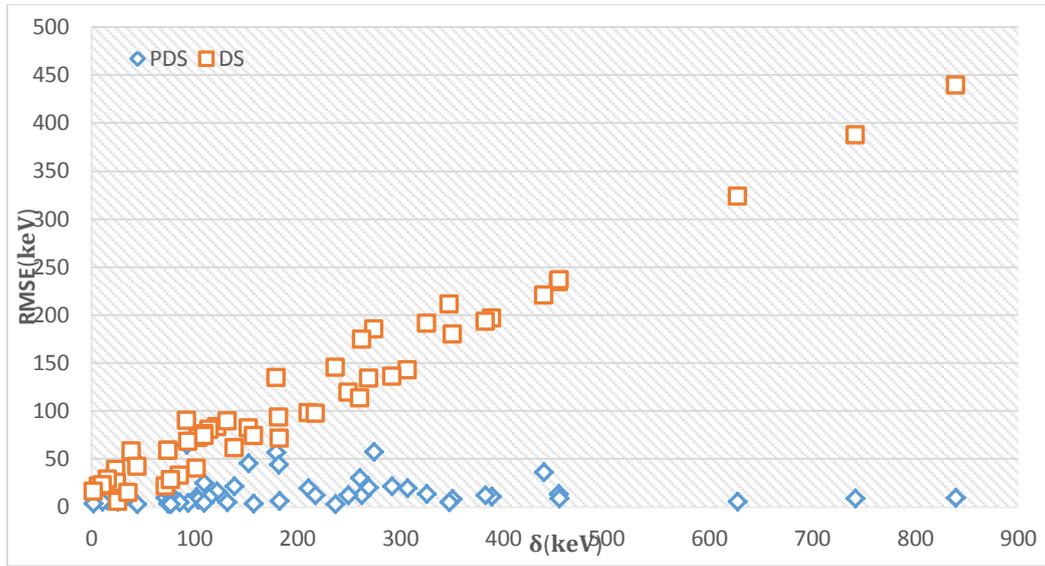

Figure 5 (color online).

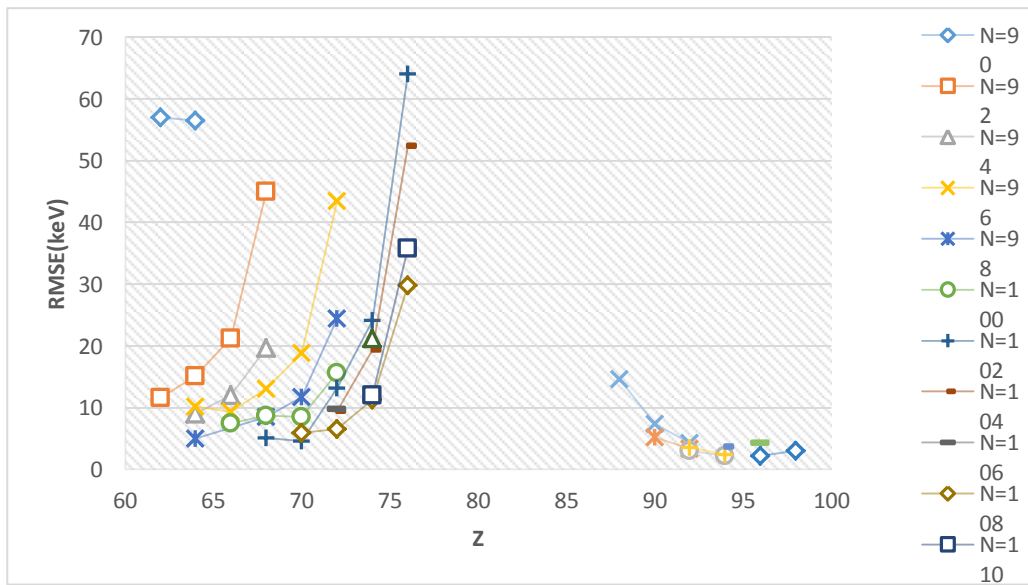

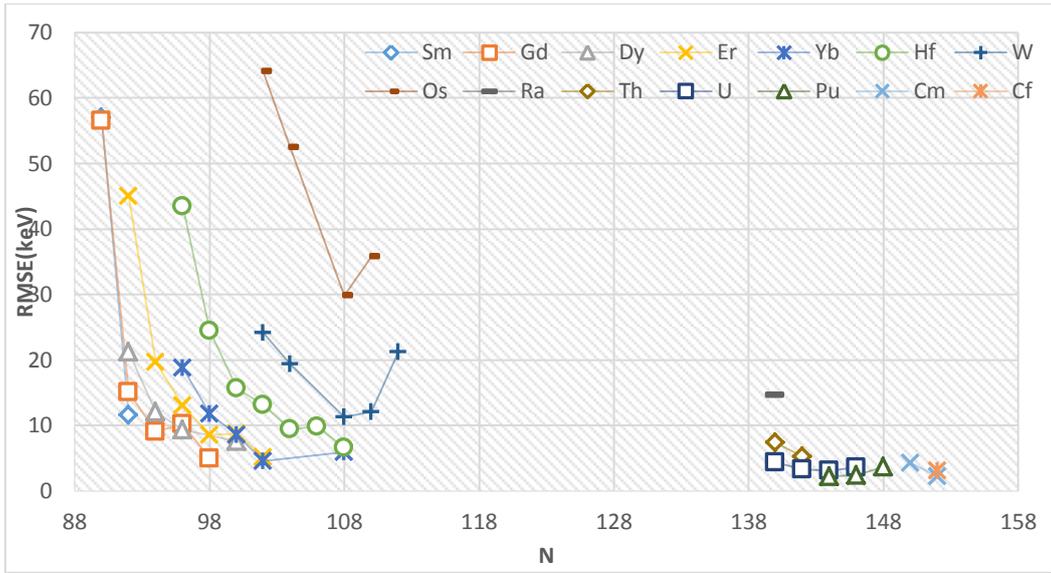

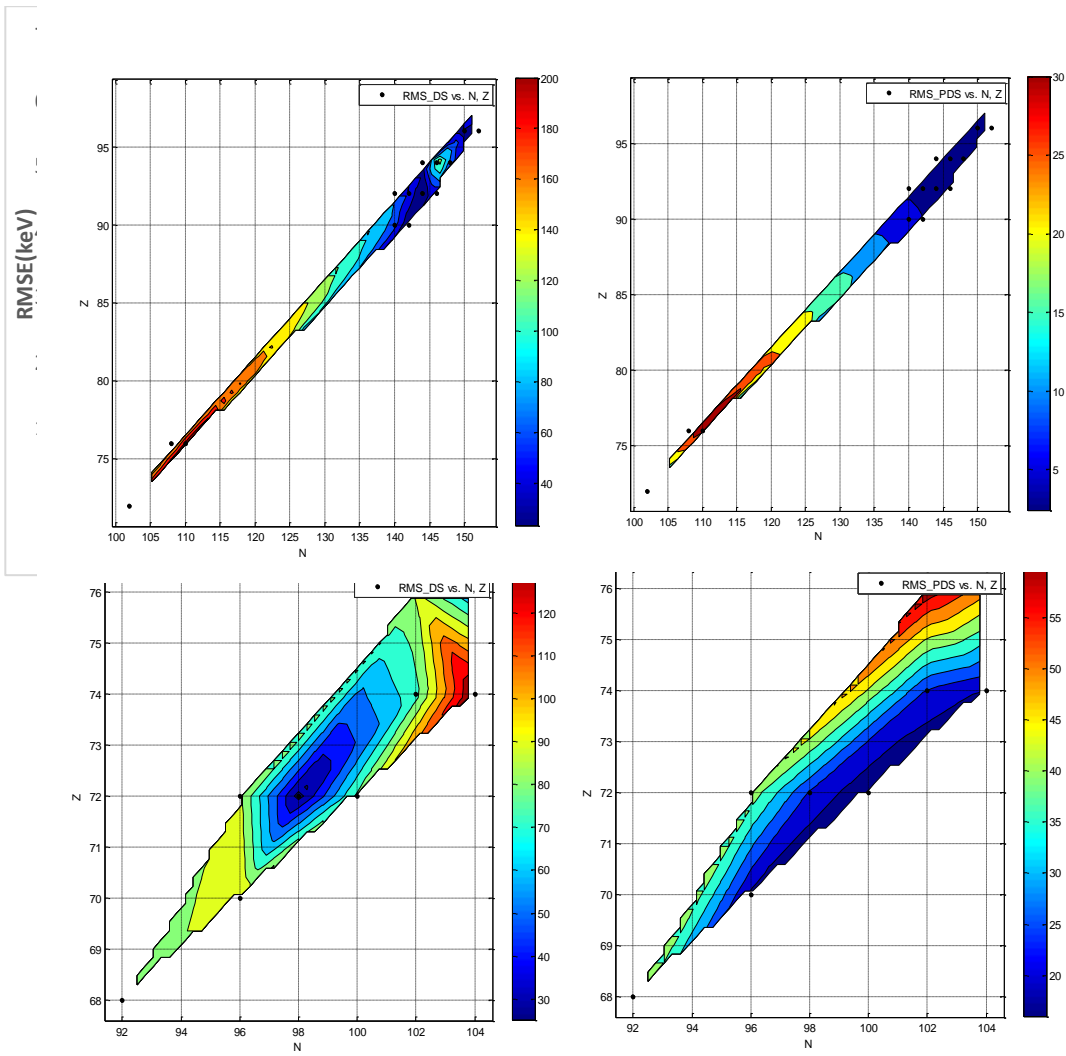